\documentclass[twocolumn,showpacs,prl]{revtex4}

\usepackage{amsmath,amssymb}
\usepackage{graphicx}

\draft

\begin{document}

\title{\large \bf Tightly Localized Stationary Pulses in Multi-Level Atomic System}

\author{Xiong-Jun Liu$^{a}$\footnote{Electronic address:
phylx@nus.edu.sg}, Xin Liu$^{b}$, Zheng-Xin Liu$^{c}$, L. C.
Kwek$^{a,d}$ and C. H. Oh$^{a}$\footnote{Electronic address:
phyohch@nus.edu.sg}} \affiliation{a. Department of Physics,
National University of Singapore,
2 Science Drive 3, Singapore 117542 \\
b. Department of Physics, Texas A\&M University, College Station,
Texas 77843-4242, USA\\
c. Theoretical Physics Division, Nankai Institute of
Mathematics,Nankai University, Tianjin 300071, P.R.China\\
d. National Institute of Education, Nanyang Technological
University, 1 Nanyang Walk, Singapore 639798}

\begin{abstract}
We show the pulse matching phenomenon can be obtained in the
general multi-level system with electromagnetically induced
transparency (EIT). For this we find a novel way to create tightly
localized stationary pulses by using counter-propagating pump
fields. The present process is a spatial compression of excitation
so that it allows us to shape and further intensify the localized
stationary pulses, without using standing waves of pump fields or
spatially modulated pump fields.

\end{abstract}

\maketitle

\indent Recently, an important progress in electromagnetically
induced transparency (EIT) \cite{EIT,exp1,6,exp2,4,5,wu} is that
experimental realization for the coherent control of stationary
pulses was achieved by using standing waves of pump fields in the
three-level system \cite{nature,three-level}. The creation of
stationary pulses can enhance the nonlinear couplings between
photons or collective excitations corresponding to stored photons,
both of which are useful for deterministic logic operations. The
key point in the creation of stationary pulses is expressed by the
pulse matching phenomenon between the forward (FW) and backward
(BW) propagating probe fields \cite{nature,three-level,four}. For
a three-level system, the technique to generate tightly localized
stationary pulses involves the use of standing waves of pump
fields with a frequency-comb or spatially modulated pump fields
\cite{three-level}. However, such tight localization is created by
a filtering process rather than a compression of excitation. Thus
the three-level technique cannot be applied directly to
applications in quantum nonlinear optics.

On the other hand, coherent manipulation of probe lights has been
studied in the four-level double $\Lambda$-type system
\cite{exp3,liu} and also in the general multi-level atomic system
that interacts with many probe and pump fields
\cite{multi-level,level2}. It has also been shown in Ref.
\cite{multi-level} that, one can convert different probe lights
into each other by manipulating the external pump fields based on
such general EIT method, indicating a sort of pulse matching
phenomenon between all applied probe fields. This observation also
motivates us to probe into a new technique of creating localized
stationary pulses based on multi-level atomic system. In this
rapid communication, we shall show the tightly localized
stationary pulses can be obtained through a spatial compression of
excitation the general multi-level EIT system.

\begin{figure}[ht]
\includegraphics[width=1.0\columnwidth]{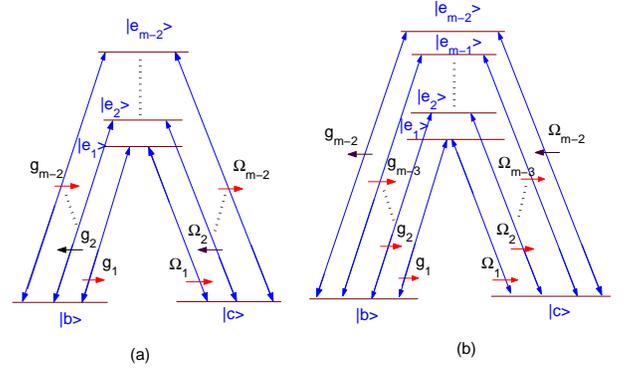}
\caption{(color online) (a) General $m$-level atomic system
coupled to $m-2$ quantized probe and classical pump fields which
propagate in $+z$ or $-z$ directions. (b) No. $1$ to No. $m-3$
pump/probe pulses propagate in the $+z$ direction, while No. $m-2$
pump/probe pulse propagates in the $-z$ direction.}\label{1}
\end{figure}

We consider the quasi-one dimensional system shown in Fig. 1(a)
for an ensemble of $m$-level atoms interacting with $m-2$
quantized probe fields which couple the transitions from the
ground state $|b\rangle$ to excited state $|e_{\sigma}\rangle$
$(1\leq \sigma\leq m-2)$ with coupling constants $g_{\sigma}$,
and $m-2$ classical pump fields which couple the transitions from
the state $|c\rangle$ to excited ones $|e_{\sigma}\rangle$ with
Rabi-frequencies $\Omega_{\sigma}(z,t)$. All probe and pump fields
are co-propagating in the $+z$ or $-z$ direction (Fig. 1(b)), and
\begin{eqnarray}\label{eqn:field1}
E_{\sigma}(z,t)&=&\sqrt{\frac{\hbar\nu_{\sigma}}{2\epsilon_0V}}\hat{\cal
E}_{\sigma}(z,t)e^{i(k_{p{\sigma}}z-\nu_{\sigma}t)},\nonumber\\
\\
\Omega_{\sigma}(z,t)&=&\Omega_{\sigma0}e^{i(k_{c{\sigma}}z-\omega_{\sigma}t)}\nonumber
\end{eqnarray}
where $\sigma=1,2,...,m-2$, $\hat{\cal E}_{\sigma}$ and
$\Omega_{{\sigma}0}$ are slowly-varying amplitudes,
$k_{p{\sigma}}$ and $k_{c{\sigma}}$, respectively $z$-component
wave vectors of probe and pump fields, can be positive or
negative. For $k_{p{\sigma}}>0$ and $k_{c{\sigma}}>0$
($k_{p{\sigma}}<0$ and $k_{c{\sigma}}<0$), it means the
${\sigma}$th pair of probe and pump fields propagate in the $+z$
($-z$) direction. We consider all transitions to be at resonance.
Under the rotating-wave approximation, the interaction
Hamiltonian can be written as:
\begin{eqnarray}\label{eqn:H1}
\hat{\mathcal V}&=&-\int\frac{dz}{L}\bigr(\hbar
N\sum_{\sigma=1}^{m-2}
g_\sigma\widetilde\sigma_{e_{\sigma}b}(z,t)\hat {\cal
E}_\sigma(z,t)+\nonumber\\
&&+\hbar N\sum_{\sigma=1}^{m-2}\Omega_{\sigma0}(t)
\widetilde\sigma_{e_{\sigma}c}(z,t)+h.c.\bigr),
\end{eqnarray}
where $N$ is the total atom number, $L$ is the length of the
medium in the $z$ direction, and the continuous atomic variables
$\widetilde\sigma_{\mu\nu}(z,t) =\frac{1}{N_z}\sum_{z_j\in
N_z}{\hat\sigma}_{\mu\nu}^j(t)$ are defined by a collection of
$N_z\gg 1$ atoms in a very small length length interval $\Delta z$
\cite{6}. $\hat\sigma_{e_{\sigma}b}^j=|e_{\sigma}^j\rangle\langle
b^j| \, {\rm e}^{-i(k_{p{\sigma}}z-\omega_{e_{\sigma}b}t)}$ and
$\hat\sigma_{e_{\sigma}c}^j=|e_{\sigma}^j\rangle\langle c^j| \,
{\rm e}^{-i(k_{c{\sigma}}z-\omega_{e_{\sigma}c}t)}$ are the
slowly-varying parts of the $j$th atomic flip operator. We note
that an essential difference between our model and the three-level
case is that for the case of multi-frequency optical pulses, here
the one- and two-photon detunings can be avoided for all optical
transitions, and no standing waves of pump fields or spatially
modulated pump fields are used.

The evolution of the slowly-varying amplitudes $\hat{\cal
E}_\sigma(z,t)$ can be described by the propagation equations
\begin{eqnarray}\label{field equation}
\left(\frac{\partial}{\partial
t}+\frac{\nu_\sigma}{k_{p\sigma}}\frac{\partial}{\partial
z}\right) \hat {\cal E}_\sigma(z,t)= ig_\sigma
N\widetilde\sigma_{be_\sigma}(z,t),
\end{eqnarray}
where we note $\nu_\sigma/k_{p\sigma}=\pm c$ for the $\pm z$
directional propagation field. Under the condition of low
excitation, i. e. $\widetilde\sigma_{bb}\approx1$, the atomic
evolution governed by the Heisenberg-Langevin equations can be
obtained by
\begin{eqnarray}\label{eqn:1}
\dot{\widetilde\sigma}_{be_\sigma}=-\gamma_{be_\sigma}
{\widetilde\sigma}_{be_\sigma} +ig_\sigma\hat{\cal
E}_\sigma+i\Omega_{\sigma0}{\widetilde\sigma}_{bc} +F_{be_\sigma},
\end{eqnarray}
\begin{eqnarray}\label{eqn:2}
\dot{\widetilde\sigma}_{bc}=
i\sum_{\sigma=1}^{m-2}\Omega_{\sigma0}{\widetilde\sigma}_{be_\sigma}
-i\sum_{\sigma=1}^{m-2}g_\sigma\hat {\cal
E}_\sigma{\widetilde\sigma}_{e_\sigma c}+F_{bc},
\end{eqnarray}
\begin{eqnarray}\label{eqn:3}
\dot{\widetilde\sigma}_{ce_\sigma}=-\gamma_{ce_\sigma}
{\widetilde\sigma}_{ce_\sigma}
+i\sum_{\sigma=1}^{m-2}g_\sigma\hat{\cal
E}_\sigma{\widetilde\sigma}_{cb} +F_{ce_\sigma},
\end{eqnarray}
where $\gamma_{\mu\nu}$ are the transversal decay rates that will
be assumed $\gamma_{be_\sigma}=\Gamma$ in the following derivation
and $F_{\mu\nu}$ are $\delta$-correlated Langevin noise operators.
From the Eq. (\ref{eqn:1}) we find in the lowest order:
${\widetilde\sigma}_{be_\sigma}=(ig_\sigma\hat{\cal
E}_\sigma+i\Omega_{\sigma0}{\widetilde\sigma}_{bc}
+F_{be_\sigma})/\Gamma$. Substitute this result into Eq.
(\ref{eqn:2}) yields $\dot{\widetilde\sigma}_{bc}=
\Gamma^{-1}\Omega_0^2{\widetilde\sigma}_{bc}
-\Gamma^{-1}\sum_{\sigma=1}^{m-2}g_\sigma\Omega_{\sigma0}\hat
{\cal E}_\sigma-i\sum_{\sigma=1}^{m-2}g_\sigma\hat {\cal
E}_\sigma{\widetilde\sigma}_{e_\sigma c}$, where
$\Omega_0=\sqrt{\sum_{\sigma=1}^{m-2}\Omega^2_{\sigma0}}$. The
Langevin noise terms are neglected in the present results, since
under the adiabatic condition the Langevin noise terms have no
effect on EIT quantum memory technique \cite{6}. For our purpose
we shall calculate $\widetilde\sigma_{bc}$ to the first order,
neglecting the small time derivatives of $\Omega_{\sigma0}$, thus
\begin{eqnarray}\label{eqn:coherence4}
\widetilde\sigma_{bc}&\approx&-\frac{1}{\Omega_0^2}\sum_{\sigma=1}^{m-2}g_\sigma\Omega_{\sigma0}\hat
{\cal E}_\sigma
-\frac{1}{\Omega_0^4}\sum_{jk\sigma}g_jg_kg_\sigma\Omega_{\sigma0}\hat{\cal
E}_j^{\dag}\hat{\cal E}_k\hat{\cal
E}_\sigma+\nonumber\\
&&+\frac{\Gamma}{\Omega_0^4}\sum_{\sigma=1}^{m-2}g_\sigma\Omega_{\sigma0}\partial_t\hat
{\cal E}_\sigma.
\end{eqnarray}
The second term in the right hand side of above equation
represents the nonlinear couplings between the probe pulses.

The dark-state polaritons (DSPs) in the general multi-level EIT
system is first obtained in \cite{multi-level}, where the
single-mode probe pulses are considered. Accordingly, the dark-
and bright-state polaritons (BSPs) in the present general
multi-level system can be defined by:
\begin{eqnarray}\label{eqn:DSP}
\hat\Psi(z,t)&=&\cos\theta\prod_{j=1}^{m-3}\cos\phi_j\hat {\cal
E}_1\nonumber\\
&&+\cos\theta\sum_{l=2}^{m-2}\sin\phi_{l-1}\prod_{j=l}^{m-3}\cos\phi_j\hat
{\cal E}_l\nonumber\\
&&-\sin\theta(t)\, \sqrt{N}\, \widetilde\sigma_{bc}(z,t),
\end{eqnarray}
\begin{eqnarray}\label{eqn:BSP}
\hat\Phi(z,t)&=&\sin\theta\prod_{j=1}^{m-3}\cos\phi_j\hat {\cal
E}_1\nonumber\\
&&+\sin\theta\sum_{l=2}^{m-2}\sin\phi_{l-1}\prod_{j=l}^{m-3}\cos\phi_j\hat
{\cal E}_l\nonumber\\
&&+\cos\theta(t)\, \sqrt{N}\, \widetilde\sigma_{bc}(z,t),
\end{eqnarray}
which are superpositions of the atomic coherence and the $m-2$
probe fields. The mixing angles $\theta$ and $\phi_j$ in the new
quantum fields are defined through
\begin{equation}\label{eqn:10}
\tan\theta=\frac{g_1g_2...g_{m-2}\sqrt{N}}{\bigr[\sum_{j=1}^{m-2}\bigr(\Omega_{j0}^2\prod_{l=1,l\neq
j}^{m-2}g_l^2\bigr)\bigr]^{1/2}}\nonumber
\end{equation}
and
\begin{equation}\label{eqn:11}
\tan\phi_j=\frac{\prod_{l=1}^{j}g_l\Omega_{j+1,0}}{\bigr[\sum_{l=1}^{j}\bigr(\Omega^2_{l0}\prod_{s=1,s\neq
l}^{j+1}g^2_s\bigr)\bigr]^{1/2}}.\nonumber
\end{equation}

Using the definitions above, one can transform the equations of
motion for the probe fields and the atomic variables into the new
field variables. With the low-excitation approximation and
neglecting the nonlinear effects we find that the DSP field
satisfies
\begin{widetext}
\begin{eqnarray}\label{eqn:DSP1}
\bigr(\partial_t +c\cos^2\theta\cos\alpha_{m-2}
\partial_z\bigr)\, \hat\Psi
&=&-\dot\theta\, \hat\Phi+\sum_{j=1}^{m-2}\dot\phi_j\cos\theta \,
\hat s_j-\frac{c}{2}\sin2\theta\cos\alpha_{m-2}\,
\partial_z\hat\Phi,
+\nonumber\\
&&+c\cos\theta\sum_{j=1}^{m-2}\prod_{l=j}^{m-3}\cos\phi_l\sin2\phi_{j-1}\bigr(\frac{1}{2c}\frac{\nu_j}{k_{pj}}
+\frac{\cos\alpha_{j-1}}{2}\bigr)\partial_z\hat s_j,
\end{eqnarray}
\end{widetext}
where we have defined
$$\cos\alpha_\sigma=c\frac{\sum_{j=1}^{\sigma}\frac{k_{pj}}{\nu_j}\Omega_{j0}^2\prod_{l=1,l\neq
j}^{\sigma}g_l^2}{\sum_{j=1}^{\sigma}\Omega_{j0}^2\prod_{l=1,l\neq
j}^{\sigma}g_l^2}, \ \sigma=1,2,...,m-3$$ and $\hat
s_j=\partial_{\phi_j}\hat\Psi/\cos\theta$. It then follows that
$\hat s_1=\prod_{j=2}^{m-3}\cos\phi_j(-\sin\phi_1{\cal
E}_1+\cos\phi_1{\cal E}_2)$, $\hat
s_2=\prod_{j=3}^{m-3}\cos\phi_j(-\sin\phi_2{(\cos\phi_1\cal
E}_1+\sin\phi_1{\cal E}_2)+\cos\phi_2{\cal E}_3)$, and generally
\begin{eqnarray}\label{eqn:s1}
\hat s_k=\prod_{j=k+1}^{m-3}\cos\phi_j\hat{\cal s}_k, \ \
k=1,2,...,m-3,
\end{eqnarray}
with
\begin{eqnarray}\label{eqn:s2}
\hat{\cal s}_k&=&\cos\phi_k{\cal
E}_{k+1}-\sin\phi_k\sum_{m=2}^k\bigr(\prod_{l=m}^{k-1}\cos\phi_l\bigr)\sin\phi_{m-1}{\cal
E}_m\nonumber\\
&-&\sin\phi_k\prod_{l=1}^{k-1}\cos\phi_l{\cal E}_1.
\end{eqnarray}
On the other hand, the equation of BSP field can be obtained as
\begin{widetext}
\begin{eqnarray}\label{eqn:BSP1}
\Phi=\frac{\Gamma}{\sqrt{N}}\biggl(\sum_{j=1}^{m-2}\bigr(\frac{\Omega_{j0}/\Omega_0}{g_j}\bigr)^2\biggl)^{1/2}
\frac{\cos\theta}{\Omega_0}\partial_t(\sin\theta\Psi-\cos\theta\Phi)+\cos\theta\bigr(g_1\Omega_{01}\sin\phi_1\hat
s_1-\sum_{l=2}^{m-3}g_l\Omega_{l0}\cos\phi_{l-1}\hat
s_{l-1}\bigr).
\end{eqnarray}
\end{widetext}
By comparing Eqs. (\ref{eqn:DSP1}) and
(\ref{eqn:BSP1}) with the corresponding DSP and BSP fields in the
three-level system, one can see a key difference is the appearance
of $\hat s_j(z,t)$ from the probe fields in our model. The
adiabatic condition in the present case can be fulfilled only if
$\hat s_j(z,t)=0$ for all $j$. However, the input probe pulses are
generally independent of each other so that the fields $\hat s_j$
need not be zero. To study the dynamics of the DSP field, we
should therefore investigate first the pulse matching between all
the probe fields needed for adiabatic condition. Bearing these
ideas in mind, we next explore the evolution of a set of normal
fields by introducing
\begin{eqnarray}\label{eqn:probe1}
\hat G_{j,j+1}=-\sin\phi_{j,j+1}\hat{\cal
E}_j(z,t)+\cos\phi_{j,j+1}\hat{\cal E}_{j+1}(z,t),
\end{eqnarray}
where $j=1, 2, ..., m-3$ and
$\tan\phi_{j,j+1}=g_j\Omega_{j+1,0}/g_{j+1}\Omega_{j0}$. From the
Eq. (\ref{field equation}) and together with the results of
${\widetilde\sigma}_{be_\sigma}$ and ${\widetilde\sigma}_{bc}$ one
can verify that the field $\hat G_{j,j+1}$ satisfies the equation
\begin{widetext}
\begin{eqnarray}\label{eqn:field3}
(\partial_t-c\cos^2\beta\cos2\phi_{j,j+1}\partial_z)\hat
G_{j,j+1}=-\frac{(g_j^2\Omega_{j+1}^2+g_{j+1}^2\Omega_j^2)N}{\Gamma}\frac{\cos^2\beta}{\Omega_0^2}\hat
G_{j,j+1}-\nonumber\\
-\frac{1}{2}g_jg_{j+1}\sqrt{N}\sin2\beta\partial_t\hat {\cal
E}_{j,j+1}+c\cos^2\beta\sin2\phi_{j,j+1}\partial_z\hat {\cal
E}_{j,j+1} +F(\hat {\cal E}_\sigma, \sigma\neq j,j+1)
\end{eqnarray}
\end{widetext} with
\begin{eqnarray}\label{eqn:mix3}
\tan^2\beta=\frac{N\Omega_j^2\Omega_{j+1}^2}{g_j^2\Omega_{j+1}^2+g_{j+1}^2\Omega_j^2}
\frac{(g_j^2-g_{j+1}^2)^2}{\Omega_0^4},\nonumber
\end{eqnarray}
and $\hat {\cal E}_{j,j+1}=\cos\phi_{j,j+1}\hat{\cal
E}_j(z,t)+\sin\phi_{j,j+1}\hat{\cal E}_{j+1}(z,t)$. $F(\hat {\cal
E}_\sigma)$ includes no $\hat {\cal E}_j$ or $\hat {\cal
E}_{j+1}$. The first term in the right hand side of Eq.
(\ref{eqn:field3}) reveals a very large absorption of $\hat
G_{j,j+1}$, which results in a large decay in the field $\hat
G_{j,j+1}$ and then the system satisfies the pulse matching
condition \cite{match,match2,multi-level}: $\hat {\cal
E}_{j+1}\rightarrow\tan\phi_{j,j+1}\hat {\cal E}_j$. For an
explicit numerical estimation, we set some typical values
\cite{exp1,exp2}: $g_j\approx g_{j+1}\sim10^5s^{-1},
N\approx10^8$, $\Gamma\approx10^8s^{-1}$, so that the life time of
field $\hat G_{j,j+1}(z,t)$ is about $\Delta t<10^{-8}s$ which is
much smaller than the storage time \cite{exp2}. Furthermore, by
introducing the adiabatic parameter
$\tau\equiv(\sum_j(1/g_j)^2)^{1/2}/\sqrt{N}T$ where $T$ is the
characteristic time scale, we can calculate the lowest order in
Eq. (\ref{eqn:BSP1}) and obtain $\hat\Phi\approx0, \hat
G_{j,j+1}\approx0$. On the other hand, under the condition of
pulse matching, one can verify that $\hat s_j(z,t)\propto\hat
G_{j,j+1}=0$. Thus equation (\ref{eqn:DSP1}) is reduced to the
shape- and state-preserving case
\begin{eqnarray}\label{eqn:DSP2}
\bigr(\partial_t +c\cos^2\theta\cos\alpha_{m-2}
\partial_z\bigr)\, \hat\Psi(z,t)=0.
\end{eqnarray}

The formula (\ref{eqn:DSP2}) is the main result of the present
work. The group velocity of the DSP field is
\begin{eqnarray}\label{eqn:group}
V_g=\cos^2\theta\frac{\sum_{j=1}^{m-2}\frac{\nu_j}{k_{pj}}\Omega_{j0}^2\prod_{l=1,l\neq
j}^{m-2}g_l^2}{\sum_{j=1}^{m-2}\Omega_{j0}^2\prod_{l=1,l\neq
j}^{m-2}g_l^2}.
\end{eqnarray}
One should bear in mind that the wave vectors $k_{pj}$ can be
positive (in the $+z$ direction) or negative (in the $-z$
direction). So, by adjusting the Rabi-frequencies for external
pump fields properly under the adiabatic condition so that
$\cos\alpha_{m-2}=0$, we can obtain a zero velocity for the DSP
field. In particular, one may set No. $1$ to No. $m-3$ pump/probe
pulses in the $+z$ direction, while No. $m-2$ pump/probe pulse in
the $-z$ direction (Fig. 1(b)) and
$\Omega_{m-2,0}=\sum_{j=1}^{m-3}\frac{g_{m-2}^2}{g_j^2}\Omega_{j0}^2$,
in an experiment so that the group velocity $V_g=0$. In this way,
we create the multi-frequency stationary pulses with each
component:
\begin{eqnarray}\label{eqn:stationary}
\hat{\cal E}_1&=&\cos\theta\prod_{j=1}^{m-3}\cos\phi_j\Psi(z,t),\nonumber\\
\hat{\cal
E}_l&=&\cos\theta\sin\phi_{l-1}\prod_{j=l}^{m-3}\cos\phi_j\Psi(z,t),\\
l&=&2,...,m-2.\nonumber
\end{eqnarray}
These components interfere to create a sharp spatial envelope. It
is helpful to present a comparison of our results with those
obtained for a three-level system \cite{three-level,nature}: i) In
the present system, all optical pulses can couple in resonance to
the corresponding atomic transitions, thus all the applied probe
fields with different frequencies contribute to the generation of
stationary pulses. This means the present process is a spatial
compression of excitation, which allows us to shape and intensify
the localized stationary pulses. Our technique can therefore be
expected to enhance further nonlinear couplings and be applied in
the most straightforward manner, e.g. to applications in quantum
nonlinear optics \cite{nonlinear}. Numerical results in Fig. 2
show how tight localization of stationary pulses can be readily
obtained when the multi-level system is applied.
\begin{figure}[ht]
\includegraphics[width=0.9\columnwidth]{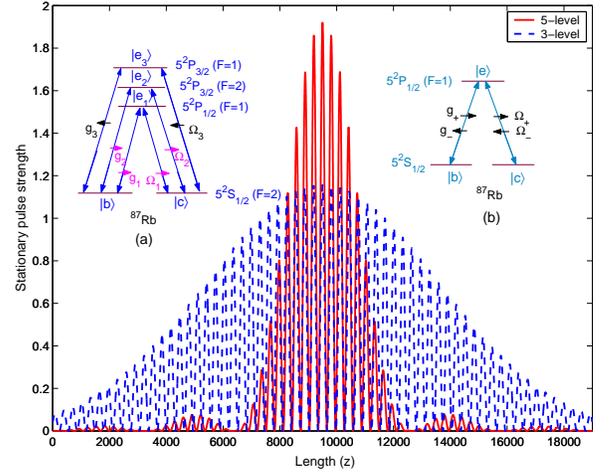}
\caption{(color online) Localization of created stationary pulses
for 5-level (red solid line) cases, where three input probe lights
are used and the parameters are set as
$\omega_{e_3e_2}=\omega_{e_2e_1}\approx\nu_2/100$ (a). As a
comparison, blue dashed line shows the stationary pulses created
in 3-level system (b) by employing one standing wave of pump
fields. The probe lights are used with the envelop
$\exp(-z^2)$.}\label{}
\end{figure}
In contrast, for a three-level system, a frequency-comb is used to
create a localized pulse, filtering the off-resonant input probe
pulses and retaining only the resonant one for creation
\cite{three-level}. Generally, the total number of probe photons
created by a frequency-comb in a three-level system is much less
than in the current model; ii) The present technique can be freely
controlled. For example, based on our results, we see that the
pulse matching in the present case is between all of the probe
pulses with different frequencies, say, $\hat {\cal
E}_{\sigma}=\prod_{j=l}^{\sigma}\tan\phi_{j,j+1}\hat {\cal E}_l \
(l\geq1, \sigma\leq{m-2})$. Thus, in principle, one can use just
one pump field to achieve the stationary pulse by adjusting its
intensity to match those of the other pump fields; iii) It
requires no standing waves in the pump fields or spatially
modulated pump fields to create localized stationary pulses.
\begin{figure}[ht]
\includegraphics[width=0.8\columnwidth]{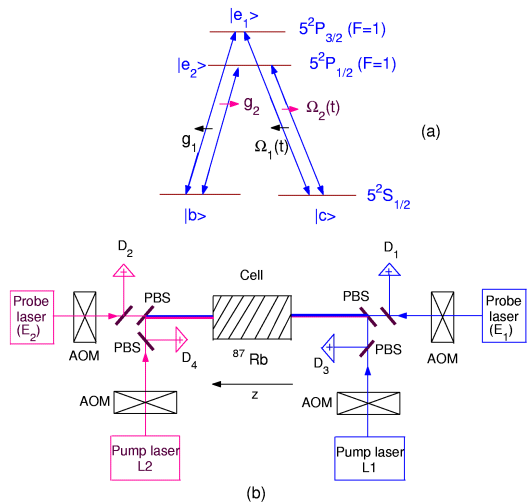}
\caption{(color online) (a)(b) Schematic of experimental
realization of stationary pulses with four-level
double-$\Lambda$-type $^{87}$Rb atoms coupled to two single-mode
quantized and two classical pump fields that propagate in $+z$ and
$-z$ directions, respectively.}\label{}
\end{figure}

Experimentally, the simplest multi-level system is an ensemble of
four-level double $\Lambda$-type $^{87}$Rb atoms. The schematic
diagram for experimental realization is shown in Fig.3. All the
atoms first are trapped in state $|b\rangle$ ($5^2S_{1/2}$) and
only the $\pm z$ directional propagation pump fields ($\Omega_1$
and $\Omega_2$) are applied to couple the transitions from
$|c\rangle$ ($5^2S_{1/2}$) to $|e_1\rangle$ ($5^2P_{1/2} (F=1)$)
and $|e_2\rangle$ ($5^2P_{3/2} (F=1)$) respectively. We then input
the probe pulses ($\hat{\cal E}_{1,2}$) and allow the system to
achieve adiabatic condition. Finally, by adjusting $\Omega_1$ or
$\Omega_2$ so that $g_1\Omega_{20}=g_2\Omega_{10}$, we can create
the stationary pulses for probe fields $\hat {\cal
E}_1(z,t)=\cos\theta\cos\phi\hat \Psi, \hat {\cal
E}_2(z,t)=\cos\theta\sin\phi\hat \Psi$, where $\hat\Psi$ is
determined by the Eq. (\ref{eqn:DSP}) with $m=4$. It can be
expected that when the level number $m$ becomes larger, the more
tightly localized stationary pulses can be created. According to
the numerical results in Fig. 2, the effect becomes substantial
when $m\geq5$.

In conclusion, we have shown the tightly localized stationary
pulses can be obtained in the general multi-level EIT system. We
have examined the dynamics of DSPs in detail and found that, all
the applied probe pulses with different frequencies contribute to
the stationary pulses. The present process is therefore a spatial
compression of excitation, which may be able to enhance further
nonlinear optical couplings and will have interesting applications
in quantum nonlinear optics \cite{nonlinear}. In particular, our
technique may open up a novel way towards the spatial compression
of many probe photons with small losses. According to the results
in \cite{multi-level}, if initially input is a non-classical probe
pulse, e.g. a quantum superposition of coherent states, we may
also generate entangled stationary pulses within our model.

\bigskip

\noindent We thank Dr. Y. Zhao (Heideberg University) for the
helpful discussions. This work is supported by NUS academic
research Grant No. WBS: R -144-000-172-101, US NSF under the grant
DMR-0547875, and by NSF of China under grants No.10275036.




\bigskip

\noindent

\end{document}